\documentclass[superscriptaddress,secnumarabic,twocolumn,
 amssymb,amsmath,nobibnotes,aps,prd,showkeys,showpacs,nofootinbib]{revtex4}
\usepackage{mathtools}
\usepackage{graphicx}
\usepackage{booktabs}
\usepackage{diagbox}
\usepackage{appendix}
\usepackage{slashbox}
\usepackage{color}
\usepackage[normalem]{ulem}
\usepackage[utf8]{inputenc}

\begin{document}

\title{Extended bodies with spin induced quadrupoles on circular equatorial orbits\\ in Kerr spacetime}
\author{Iason Timogiannis}
\email{timogian@phys.uoa.gr}
\affiliation{Section of Astrophysics, Astronomy, and Mechanics, Department of Physics,
University of Athens, Panepistimiopolis Zografos GR15783, Athens, Greece}
\author{Georgios Lukes-Gerakopoulos}
\email{gglukes@gmail.com}
\affiliation{Astronomical Institute of the Czech Academy of Sciences,
Bo\v{c}n\'{i} II 1401/1a, CZ-141 00 Prague, Czech Republic}
\author{Theocharis A. Apostolatos}
\email{thapostol@phys.uoa.gr}
\affiliation{Section of Astrophysics, Astronomy, and Mechanics, Department of Physics,
University of Athens, Panepistimiopolis Zografos GR15783, Athens, Greece}

\begin{abstract}
This work discusses the motion of extended test bodies as described by the Mathisson-Papapetrou-Dixon (MPD) equations in the pole-dipole-quadrupole approximation. We focus on the case that the quadrupole is solely induced by the spin of the body which is assumed to move on a circular equatorial orbit in a Kerr background. To fix the center of mass of the MPD body we use two different spin supplementary conditions (SSCs): the Tulczyjew-Dixon SSC and the Mathisson-Pirani SSC. We provide the frequencies of the circular equatorial orbits for the pole-dipole-(spin induced) quadrupole approximation of the body for both SSCs. In the process we develop an explicit four-velocity four-momentum relation for a pole-dipole-quadrupole body under the Mathisson-Pirani SSC.
\end{abstract}

\pacs{~~}
\maketitle

\section{Introduction}

The discussion of finite size effects of a body moving in a curved background dates back at least to the pioneering work of Mathisson \cite{Mathisson37}. Mathisson introduced a ``gravitational skeleton" for the extended body, i.e., a multipole expansion of the body's energy-momentum tensor $T^{\mu\nu}$ around a reference point $z^\mu$. This point is called centroid and corresponds to the center of mass of the body. In practice, the multipole expansion is truncated at certain order, which in our study is the quadrupole, and all the higher multipoles are ignored. The motion of this test body in a curved spacetime is described by the so called Mathisson-Papapetrou-Dixon (MPD) equations \cite{Mathisson37,Papapetrou51,Dixon74b}. When only gravitational interactions are considered, the MPD equations in the pole-dipole-quadrupole approximation read
\begin{eqnarray}
\dot{p}^{\mu}& = & -\frac{1}{2}R^{\mu}_{~\nu \alpha \beta}u^{\nu} S^{\alpha \beta}+F^\mu, \label{eq:MPD_P}  \\
\dot{S}^{\mu \nu} & =& 2 p^{[\mu} u^{\nu]}+F^{\mu\nu},\label{eq:MPD_S}
\end{eqnarray}
where $R^{\mu}_{~\nu \alpha \beta}$ is the Riemann tensor of the background gravitational field, $p^\mu$ is the four-momentum of the test body, $\dot{~}=u_\mu \nabla^{\mu}$ denotes its covariant derivative along the four-velocity $u^\mu$ and $S^{\mu\nu}$ represents the spin tensor of the body. The quadrupole contribution to the MPD equations comes only from the terms  $F^\mu=-\dfrac{1}{6}J^{\alpha \beta \gamma \delta} \nabla^{\mu}R_{\alpha \beta \gamma \delta}$ and $F^{\mu\nu}=\dfrac{4}{3} J^{\alpha \beta \gamma [\mu}R^{\nu]}_{~\gamma \alpha \beta }$, with $J^{\alpha \beta \gamma \delta}$ designating the quadrupole tensor. There is no corresponding evolution equation for the quadrupole tensor, which is simply defined by the matter structure of the body \cite{Steinhoff:2012rw,Bini:2013uwa}.

The MPD equations are an underdetermined set of evolution equations needing in general four additional closing relations \cite{LG17}. By choosing $u^\mu$ to represent the four-velocity of the body, i.e., by selecting the evolution parameter of the MPD equations to be the proper time, we have introduced the constraint $u^\mu u_\mu=-1$. The other three constraints\footnote{Only three of the four equations of a SSC described by Eq.~\eqref{eq:SSC} are linearly independent.} originate from the imposed spin supplementary condition (SSC), which fixes the center of the mass of the body. A SSC is given by the general relation
\begin{align}\label{eq:SSC}
    V^\mu S_{\mu\nu}=0,
\end{align}
where $V^\mu$ is a future oriented time-like vector, which is often accompanied by the normalization condition $V^\mu V_\mu=-1$. 

Having defined the reference vector $V^\mu$, a four-vector of the spin can be defined as
\begin{align}
\label{eq:SpinVect}
 S_\mu := -\frac{1}{2} \epsilon_{\mu\nu\rho\sigma}
          \, V^\nu \, S^{\rho\sigma} \; ,
\end{align}
while the inverse relation of Eq.~\eqref{eq:SpinVect} reads
\begin{align} \label{eq:SpinTensor}
    S^{\rho\sigma}=-\epsilon^{\rho\sigma\nu\kappa}S_\nu V_\kappa,
\end{align}
where $\epsilon_{\mu\nu\rho\sigma}$ is the Levi-Civita tensor. The latter relation after some calculations entails that
\begin{align}\label{eq:SpinTSpinTC}
    S^{\mu\nu} S_{\nu\beta}=S^\mu S_\beta-S^2 {h^\mu}_\beta,
\end{align}
where ${h^\mu}_\beta={\delta^\mu}_\beta+V^\mu V_\beta$ is the space projector orthogonal to $V^\mu$ and $S^2=\dfrac{1}{2}S^{\mu\nu}S_{\mu\nu}=S^\mu S_\mu$ is the square of the magnitude of the spin. Moreover, with the help of the reference vector the quadrupole moment can be decomposed in the following manner \cite{Harte20} 
\begin{equation}\label{eq:QuadTens}
J^{\alpha \beta \gamma \delta}=\tau^{\alpha \beta \gamma \delta}-3 V^{[\alpha}Q^{\beta][\gamma}V^{\delta]}-V^{[\alpha}\Pi^{\beta] \gamma \delta}-V^{[\gamma}\Pi^{\delta] \alpha \beta},
\end{equation}
where $Q^{\mu\nu}$ is the mass quadrupole, $\Pi^{\mu\nu\kappa}$ is the flow (or current)
quadrupole and $\tau^{\alpha \beta \gamma \delta}$ is the stress quadrupole \cite{Ehlers1977}. 

In this work we focus on the spin induced quadrupole model, for which the mass quadrupole reads
\begin{align}\label{eq:SpinInd}
    Q^{\beta\gamma}=c_{S^2}{S^\beta}_\alpha S^{\alpha\gamma},
\end{align}
where $c_{S^2}$ is a constant depending on the internal structure of the body \cite{Steinhoff:2012rw}. Note that due to Eq.~\eqref{eq:SpinTSpinTC} the right hand side can be rewritten in terms of the spin four-vector. We are going to exploit this observation in our investigation of circular equatorial orbits (CEOs). 

Studies of CEOs are an important first step in the comprehension of the orbital dynamics around a black hole \cite{Bardeen72,Rasband73,Tod76,Steinhoff:2012rw,Bini15,Harms16,Timogiannis21,Timogiannis22}. This fact is especially pronounced in the modelling of extreme mass ratio inspirals  \cite{Barack05,Barack07,Yunes10,Isoyama14,Piovano20,Pound20,Mathews22}, in which a stellar compact object, like a black hole or a neutron star, inspirals in the background of a supermassive black hole. Even when calibrating the GW waveforms, the starting points are CEOs \cite{Albertini22,Albertini22b}. Hence, in our study we focus on CEOs of an extended test body in pole-dipole-(spin induced) quadrupole approximation around a Kerr black hole.  

Our study on this topic differs from previous ones \cite{Steinhoff:2012rw,Bini15} in several ways. We do not consider only the Tulzcyjew-Dixon (TD) SSC \cite{Tulczyjew59,Dixon64}
\begin{align} \label{eq:TDSSC}
    p^\mu S_{\mu\nu}=0,
\end{align}
but also the Mathisson-Pirani (MP) one \cite{Mathisson37,Pirani56}
\begin{align}\label{eq:MPSSC}
    u^\mu S_{\mu\nu}=0.
\end{align}
We do not use the effective potential approach to find CEOs for the TD SSC, but an analytic treatment introduced in Ref.~\cite{Khodagholizadeh20} and later further developed in our works \cite{Timogiannis21}. This treatment is exact in the framework of the pole-dipole-(spin induced) quadrupole approximation and does not employ any approximation in powers of the spin magnitude $S$.

The rest of the article is organized as follows.  Sec.~\ref{sec:UP} briefs the four-velocity four-momentum relation for a pole-dipole-quadrupole body approximation under the TD SSC presented in Ref.~\cite{Steinhoff:2012rw} and introduces such a relation under the MP SSC as well. Sec.~\ref{sec:CEO} is divided into two main parts; the first subsection provides a novel procedure on how to find CEOs for a pole-dipole-(spin induced) quadrupole body under the TD SSC, while the second subsection provides the CEOs for such a body under the MP SSC for the first time. Finally, Sec.~\ref{sec:Conc} summarizes the findings.

\textit{Units and notation:} We use geometric units, in which the speed of light and the gravitational constant are normalized to  $c=G=1$. The Riemann tensor is defined as $R^\mu_{\nu\kappa\lambda}=\Gamma^\mu _{\kappa \alpha} \Gamma^\alpha _{\lambda \nu}-\partial_\lambda \Gamma^\mu _{\kappa \nu}-\Gamma^\mu _{\lambda\alpha}\Gamma^\alpha _{\kappa \nu}+\partial_\kappa \Gamma^\mu _{\lambda \nu}$, while the Christoffel symbols are computed from the metric with signature $(-,+,+,+)$. Greek indices run from 0 to 3. The Levi-Civita tensor is given by $\epsilon_{\mu\nu\rho\sigma}=\sqrt{-g}\,\tilde{\epsilon}_{\mu\nu\rho\sigma}$, where $\tilde{\epsilon}_{\mu\nu\rho\sigma}$ is the Levi-Civita symbol and $g$ is the determinant of the background metric. In the present paper the background metric coincides with the Kerr black hole metric. The central black hole's mass is denoted by $M$, while for the test body we employ the two notions of mass $\mu:=\sqrt{-p^\mu p_\mu}$ and $m:=-u^\mu p_\mu$. To be consistent with our definition of the reference vector, for the TD SSC we use $V^\mu=p^\mu/\mu$. Dimensionless quantities are denoted by the symbol $\hat{~}$. For instance the dimensionless Kerr parameter $a$ is denoted as $\hat{a}=a/M$ and the dimensionless radius $r$  is similarly denoted as $\hat{r}=r/M$. However, for the dimensionless measure of the spin we follow a different convention depending on the SSC:
for the TD SSC the dimensionless spin measure is denoted as $\sigma=\dfrac{S}{\mu M}$, while for the MP SSC we use $\sigma=\dfrac{S}{m M}$. In a similar fashion, we set $\hat{c}_{S^2}=\mu c_{S^2}$ for the TD SSC and $\hat{c}_{S^2}=mc_{S^2}$ for the MP SSC.

\section{Velocity-Momentum relations}\label{sec:UP}

Even if the MPD equation system closes by choosing a SSC and an evolution parameter, it is useful to have a relation between the four-velocity and the four-momentum in order to evolve the MPD equations numerically. Such relations have been provided for the TD SSC in the pole-dipole case \cite{Ehlers1977} and in the pole-dipole-quadrupole case \cite{Steinhoff:2012rw}, while for the MP SSC the pole-dipole got its velocity momentum relation recently in Ref.~\cite{Costa18}. In this Section we review the existence of a $u^\nu=f(p^\nu)$ relation by taking advantage of the analytical framework set up in \cite{Steinhoff:2012rw} and \cite{Costa18}. Below, we treat each SSC separately. 

\subsection{TD SSC}
For the Tulczyjew-Dixon SSC Steinhoff and Puetzfeld derived in \cite{Steinhoff:2012rw} a momentum-velocity relation, analogous to the pole-dipole case, by defining the quantity
\begin{equation}
    \tilde{p}^\nu=\frac{1}{\mu^2}\biggl(m p^\nu+F_\kappa S^{\nu\kappa}-p_\kappa F^{\nu\kappa}\biggr).
\end{equation}
The desired relation reads
\begin{equation} \label{eq:Steinh}
    u^\nu=\tilde{p}^\nu+\frac{2 R_{\kappa\lambda\mu\sigma}S^{\nu\kappa}S^{\mu\sigma}\tilde{p}^\lambda}{4\mu^2+R_{\alpha\beta\gamma\delta}S^{\alpha\beta}S^{\gamma\delta}}.
\end{equation}
Note that for the pole-dipole-quadrupole approximation in general $S_\nu u^\nu \neq 0$. To prove this statement one starts by contracting the evolution equation of the spin tensor Eq.~\eqref{eq:MPD_S} with the covariant four-momentum $p_\mu$ in order to get
\begin{equation} \label{eq:momvelTD1}
    p_\mu \dot{S}^{\mu\nu}=-\mu^2 u^\nu+m p^\nu+p_\mu F^{\mu\nu}.
\end{equation}
By virtue of Eq.~\eqref{eq:TDSSC} $p_\mu \dot{S}^{\mu\nu}=-\dot{p}_\mu S^{\mu\nu}$, while the contraction of Eq.~\eqref{eq:momvelTD1} with the spin $S_\nu$, finally yields
\begin{equation}
S_\nu u^\nu=\frac{p_\mu S_\nu F^{\mu\nu}}{\mu^2}.  
\end{equation}
As a result, the quadrupole contribution implies that while $S_\nu p^\nu=0$, $S_\nu u^\nu\neq0$, contrarily to the pole-dipole approximation where
\begin{equation} \label{eq:isotita}
 S_\nu p^\nu=S_\nu u^\nu=0,  
\end{equation}
under the TD SSC. In Sec.~\ref{sec:CEO} we shall see that the demand for CEOs, along with its associated assumptions for the spin induced quadrupole, eventually render Eq.~\eqref{eq:isotita} valid. 

Notice that Eq.~\eqref{eq:Steinh} combined with the normalization condition of the four-velocity, i.e. $u_\nu u^\nu=-1$, implies a relation between $\mu$ (dynamical mass) and $m$ (kinematical mass). The aforementioned expression reads
\begin{equation} \label{eq:massquadr}
    m^2 \biggl(\frac{\mathcal{F}_4}{\mathcal{F}_0^2}-\frac{1}{\mu^2}\biggr)+m\biggl(\frac{\mathcal{F}_2}{\mathcal{F}_0}+\frac{\mathcal{F}_5}{\mathcal{F}_0^2}\biggr)+\biggl(\mathcal{F}_1+\frac{\mathcal{F}_3}{\mathcal{F}_0}+\frac{\mathcal{F}_6}{\mathcal{F}_0^2}+1\biggr)=0,
\end{equation}
with the corresponding coefficients given by
\begin{align*}
    \mathcal{F}_0&=4\mu^2+R_{\alpha\beta\gamma\delta}S^{\alpha\beta}S^{\gamma\delta},\\
    \mu^4 \mathcal{F}_1&=F_\kappa S_{\mu\alpha}F^\alpha S^{\mu\kappa}-2F_\kappa F_{\mu\nu}p^\nu S^{\mu\kappa}+F_{\mu\lambda}p_\nu p^\lambda F^{\mu\nu},\\
    \mu^2 \mathcal{F}_2&=4 R_{\kappa\lambda\nu\sigma}S^{\mu\kappa}S^{\nu\sigma}p^\lambda(S_{\mu\alpha}F^\alpha-F_{\mu\beta}p^\beta), \\
    \mu^2\mathcal{F}_3&=4 R_{\kappa\lambda\nu\sigma}S^{\mu\kappa}S^{\nu\sigma}(F_\gamma S^{\lambda\gamma}-p_\delta F^{\lambda\delta})(S_{\mu\alpha}F^\alpha-F_{\mu\beta}p^\beta),\\
    \mu^4\mathcal{F}_4&=4 R_{\kappa\lambda\nu\sigma}S^{\mu\kappa}S^{\nu\sigma}R^\pi_{~\rho\epsilon\zeta}S_{\mu\pi}S^{\epsilon\zeta}p^\lambda p^\rho,\\
    \mu^4 \mathcal{F}_5&=4 R_{\kappa\lambda\nu\sigma}S^{\mu\kappa}S^{\nu\sigma}R^\pi_{~\rho\epsilon\zeta}S_{\mu\pi}S^{\epsilon\zeta}p^\lambda(F_\alpha S^{\rho\alpha}-p_\beta F^{\rho\beta})\\
    &+4 R_{\kappa\lambda\nu\sigma}S^{\mu\kappa}S^{\nu\sigma}R^\pi_{~\rho\epsilon\zeta}S_{\mu\pi}S^{\epsilon\zeta}p^\rho(F_\xi S^{\lambda\xi}-p_\tau F^{\lambda\tau}),\\
    \mu^4\mathcal{F}_6&=4 R_{\kappa\lambda\nu\sigma}S^{\mu\kappa}S^{\nu\sigma}R^\pi_{~\rho\epsilon\zeta}S_{\mu\pi}S^{\epsilon\zeta}F_\alpha S^{\rho\alpha}(F_\xi S^{\lambda\xi}-p_\tau F^{\lambda\tau})\\
    &-4 R_{\kappa\lambda\nu\sigma}S^{\mu\kappa}S^{\nu\sigma}R^\pi_{~\rho\epsilon\zeta}S_{\mu\pi}S^{\epsilon\zeta}p_\beta F^{\rho\beta}(F_\xi S^{\lambda\xi}-p_\tau F^{\lambda\tau}).
    \end{align*}
    For a brief crosscheck notice that Eq.~\eqref{eq:massquadr} reduces to the expression given in \cite{Witzany}, when $J^{\alpha\beta\gamma\delta}=0$. In general, Eq.~\eqref{eq:massquadr} has two distinct roots $m_+$, $m_-$. The physically accepted solution however, is $m_+$, since it is the respective positive one in the pole-dipole limit.

\subsection{MP SSC} \label{sec:MPSSC}

The derivation of an explicit momentum-velocity relation under the MP closure choice, even in the pole-dipole approximation, has been missing for almost a century. This rendered numerical calculations, for instance,   more cumbersome than for the TD SSC choice. It was not until recently that Costa et al.~\cite{Costa18} shed light on the matter and extracted the long sought relation for the pole-dipole approximation. In this part of the article we attempt to derive an analogous relation for the pole-dipole-quadrupole approximation, based on fundamental principles. In addition, in Appendix~\ref{sec:app1} we present an alternative way of treating the problem, by taking into account the framework employed in \cite{Costa18} for the pole-dipole case, that leads to the same result as given in this section.

Our starting point is Eq.~\eqref{eq:SpinTSpinTC}, implemented in the MP reference frame, which yields
\begin{equation} \label{eq:SpintensorsMP}
    S^{\mu\nu}S_{\nu\beta}=S^\mu S_\beta-S^2 (\delta^\mu_\beta+u^\mu u_\beta).
\end{equation}
The contraction of Eq.~\eqref{eq:SpintensorsMP} with the four-momentum $p^\beta$ gives
\begin{equation} \label{eq:momvelMP}
   m u^\nu=p^\nu+\dfrac{S_{\mu\beta}p^\beta S^{\nu\mu}-S_\beta p^\beta S^\nu}{S^2}, 
\end{equation}
which is the desirable result. We wish to underline that in the pole-dipole case, where $S_\beta p^\beta=0$, the last term vanishes and Eq.~\eqref{eq:momvelMP} is identical to the expression in \cite{Costa18}. The last term in Eq.~\eqref{eq:momvelMP} pertains to the quadrupole tensor through $F^{\mu\nu}$. To confirm this, one contracts the evolution equation of the spin tensor, Eq.~\eqref{eq:MPD_S}, with the covariant four-velocity $u_\nu$ in order to get
\begin{equation} \label{eq:Mpmomvel}
 u_\nu \dot{S}^{\mu\nu}=-p^\mu+m u^\mu+u_\nu F^{\mu\nu}.   
\end{equation}
The adoption of MP SSC yields $u_\nu \dot{S}^{\mu\nu}=-\dot{u}_\nu S^{\mu\nu}$ and the contraction of Eq.~\eqref{eq:Mpmomvel} with $S_\mu$ reveals
\begin{equation} \label{eq:important}
    S_\mu p^\mu=u_\nu S_\mu F^{\mu\nu}\, .
\end{equation}
Furthermore, if we contract Eq.~\eqref{eq:momvelMP} with the covariant component of the four-momentum $p_\nu$, we conclude that
\begin{equation}
m^2=\mu^2-\frac{1}{S^2}\biggl[S_{\mu\beta}p_\nu p^\beta S^{\nu\mu}-(p_\nu S^\nu)^2\biggr].    
\end{equation}

Even if in the RHS of Eq.~\eqref{eq:momvelMP} $u^\mu$ does not appear explicitly, Eq.~\eqref{eq:momvelMP} is not on the same footing as its pole-dipole counterpart, due to the emergence of $S^\mu$ on the RHS. Namely, the RHS of Eq.~\eqref{eq:momvelMP} should be only function of $p^\mu$ and $S^{\mu\nu}$; the  $S^\mu$ presence indirectly implies the existence of $u^\mu$. Despite this shortcoming, we found it very useful in our study of CEOs discussed in the next section.

\section{CIRCULAR EQUATORIAL ORBITS} \label{sec:CEO}

The study of an extended spinning body moving on a CEO in a fixed background consists of selecting the appropriate initial data for the variables $\{z^\nu, p^\nu, S^{\mu\nu}\}$, so that circular equatorial motion is obtained upon evolution of the MPD equations. The vast majority of attempts in the literature include effective potential methods \cite{Tod76,Steinhoff:2012rw,Bini15,Harms16}. In our approach we take advantage of an algorithm introduced in Ref.~\cite{Khodagholizadeh20} and improved in \cite{Timogiannis21, Timogiannis22} for the pole-dipole approximation. We work in Kerr spacetime using Boyer-Lindquist coordinates, where the metric components are functions of $r$ and $\theta$, and more precisely
\begin{align*}
    g_{tt}&=-1+\dfrac{2 M r}{\Sigma}, &  g_{t\phi}&=-\dfrac{2 a M r \sin^2\theta}{\Sigma}, & g_{\theta\theta}=\Sigma, \\ 
    g_{\phi\phi}&=\dfrac{\Lambda \sin^2\theta}{\Sigma}, &  g_{r r}&=\dfrac{\Sigma}{\Delta}, 
\end{align*}
with the parameters
\begin{align*}
    \Sigma&=r^2+a^2 \cos^2 \theta, \\
    \Lambda&=\varpi^4-a^2 \Delta \sin^2 \theta, \\
    \Delta&=\varpi^2-2 M r, \\
    \varpi&=\sqrt{r^2+a^2}.
\end{align*}
 The present study is restricted in the equatorial plane, with the spatial coordinates described by $r=constant$, $\theta=\pi/2$ and $\phi=\Omega t$. The orbital frequency of the extended, spinning body is $\Omega=\dfrac{u^\phi}{u^t}$, whereas the radial and polar components of the four-velocity are set to $u^r=0$ and $u^\theta=0$. Under these assumptions the normalization condition of the four-velocity, i.e., $u^\nu u_\nu=-1$, entails that
 \begin{equation} \label{eq:OrbFre}
   u^t=\frac{1}{\sqrt{-g_{tt}-2g_{t\phi}\Omega-g_{\phi\phi}\Omega^2}}.  
 \end{equation}
In addition, we demand that the spin four-vector, $S^\nu= S^\theta\delta^\nu_\theta$ is aligned (or anti-aligned) with the angular momentum $J_z$, which by convention is always pointing along the positive $z$-direction in our setup. Subsequently, Eq.~\eqref{eq:SpinTensor} manifestly shows that the spin tensor is characterized by four non vanishing components, for both closure choices examined here
\begin{align} 
    S^{tr}&=-S^{rt}=-S\sqrt{-\frac{g_{\theta\theta}}{g}}V_\phi, \label{eq:s1}\\
    S^{r\phi}&=-S^{\phi r}=-S\sqrt{-\frac{g_{\theta\theta}}{g}}V_t.
\end{align}
 
 Under the restrictions imposed by the introduced setup, the set of the MPD Eqs.~\eqref{eq:MPD_P}, \eqref{eq:MPD_S} leads to trivial identities, apart from the $\dot{p}^{r}$ and $\dot{S}^{t\phi}$ components. Note that this behaviour was also present in the pole-dipole case and was outlined in \cite{Khodagholizadeh20, Timogiannis21, Timogiannis22}.     

\subsection{CEOS UNDER TD SSC} \label{sec:CEOTD}

\begin{table}[h]
\centering
\begin{tabular}{|c|ccc|ccc|}
\hline
\multicolumn{7}{|c|}{$\hat{c}_{S^2}=1$}\\
\hline
&\multicolumn{3}{|c|}{$\hat{r}_{\rm ISCO}$}
&\multicolumn{3}{|c|}{$M \Omega_{\rm ISCO}$}\\
\hline
\backslashbox{${\hat a}$}{\\ ${\sigma}$}&$-0.1$ &0 &0.1 &$-0.1$ &0 &0.1 \\
\hline
0 & 6.1643 & 6 & 5.8377 & 0.06596 &  0.06804 & 0.07012\\
0.1 & 5.8313& 5.6693& 5.5094& 0.07120&  0.07354& 0.07586\\
0.3 & 5.1351& 4.9786& 4.8248& 0.08466& 0.08765& 0.09063\\
0.5 & 4.3824& 4.2330& 4.0876& 0.10454& 0.10859& 0.11258\\
0.7 & 3.5259& 3.3931& 3.3002& 0.13829& 0.14388& 0.14724\\
0.9 & 2.4018& 2.3209& 2.2427& 0.21925& 0.22544& 0.23152\\
\hline
\multicolumn{7}{|c|}{$\hat{c}_{S^2}=6$}\\
\hline
&\multicolumn{3}{|c|}{$\hat{r}_{\rm ISCO}$}
&\multicolumn{3}{|c|}{$M \Omega_{\rm ISCO}$}\\
\hline
\backslashbox{${\hat a}$}{\\ ${\sigma}$}&$-0.1$ &0 &0.1 &$-0.1$ &0 &0.1 \\
\hline
0 & 6.1893 & 6 & 5.8627 & 0.06553& 0.06804 & 0.06965\\
0.1  & 5.8565& 5.6693& 5.5346& 0.07072&  0.07354& 0.07532\\
0.3  & 5.1604& 4.9786& 4.8501& 0.08400& 0.08765& 0.08990\\
0.5  & 4.4065& 4.2330& 4.1122& 0.10366& 0.10859& 0.11155\\
0.7  & 3.5438& 3.3931& 3.2815& 0.13720&  0.14388& 0.14818\\
0.9  & 2.4094& 2.3209& 2.2504& 0.21801&  0.22544& 0.23011\\
\hline
\end{tabular}
\caption{Dimensionless innermost stable circular orbit radius along with its associated frequency for an extended spinning body governed by the TD SSC, for two special cases of $\hat{c}_{S^2}$. 
}
\label{tab:ISCOBini}
\end{table}

\begin{table}[h]
\centering
\normalsize
\begin{tabular}{|c|c|c|c|}
\hline
\multicolumn{4}{|c|}{$\hat{c}_{S^2}=1$}\\
\hline
\backslashbox{$\sigma$}{\\ $\hat{a}$}&$-0.9$ &0 &0.9\\
\hline
$-0.9$ & 0.03390 & 0.03260& 0.03135\\
$-0.1$ & 0.03275& 0.03177& 0.03085\\
$+0.1$ & 0.03234& 0.03147& 0.03064\\
$+0.9$ & 0.03027& 0.02993& 0.02956\\
\hline
\multicolumn{4}{|c|}{$\hat{c}_{S^2}=6$}\\
\hline
\backslashbox{$\sigma$}{\\  ${\hat a}$}&$-0.9$ &0 &0.9\\
\hline
$-0.9$ & 0.03105 & 0.03038 & 0.02954\\
$-0.1$  & 0.03271& 0.03174& 0.03082\\
$+0.1$  & 0.03231& 0.03144& 0.03062\\
$+0.9$  & 0.02766& 0.02790& 0.02790\\
\hline
\end{tabular}
\caption{ The Table depicts the orbital frequencies $\hat{\Omega}$ of a spinning body (black hole or neutron star) moving in a corrotating/counterotating circular equatorial orbit of $\hat{r}=10$, under the TD SSC.    
}
\label{tab:CEOTD}
\end{table}

For the TD SSC the reference four-vector has to be replaced by $V^\nu:=p^\nu/\mu$. The evolution equation of the dynamical rest mass is described by \cite{Steinhoff:2012rw}

\begin{equation} \label{eq:mudot}
    \dot{\mu}=\dfrac{\mu \dot{R}_{\rho \beta \gamma \delta}J^{\rho \beta \gamma \delta}}{6m}-\dfrac{4\dot{p}_\alpha p_\beta R^{[\alpha}_{~\gamma \pi \kappa } J^{\beta]\gamma \pi \kappa }} {3\mu m} ,
\end{equation}
while the covariant derivative of the spin length reads \cite{Steinhoff:2012rw}

\begin{equation} \label{eq:Sdot}
    S \dot{S}= \dfrac{2 S_{\alpha \beta} R^\alpha_{~\rho \kappa \lambda} J^{\beta \rho \kappa \lambda}}{3}.
\end{equation}
The difference in the signs of Eqs.~\eqref{eq:mudot}, \eqref{eq:Sdot} compared to the corresponding relations found in \cite{Steinhoff:2012rw} originates from the different signature convention of the metric tensor. The demand for CEOs combined with the imposed TD SSC, renders both $\mu$ and $S$ conserved quantities, i.e., $\dot{\mu}=0$, $\dot{S}=0$. The latter is not generally true within the pole-dipole-quadrupole regime. In addition, Eq.~\eqref{eq:Steinh} implies that $u^r \propto p^r$ as well as $u^\theta \propto p^\theta$, which are features that also hold in the pole-dipole case.

For the non vanishing components of the MPD Eqs.~\eqref{eq:MPD_P}, \eqref{eq:MPD_S} one has
\begin{align} 
    &3c_{S^2} M S^2 [r^2 \mu^2+5 a^2 (p^t)^2-2a (5a^2+4r^2)p^t p^\phi  \nonumber \\ 
    &+(5a^4+8a^2 r^2+3 r^4)(p^\phi)^2]=  \nonumber \\ 
    &2 \mu M r^2 p^t [-(3 a S+\mu r^2)u^t+(3 a^2 S+r^2 S+a \mu r^2)u^\phi] \nonumber \\  &+ 2 \mu r^2 p^\phi [M(3 a^2 S+2 r^2 S+a \mu r^2)u^t \nonumber \\ 
    &-(3 a^3 M S+3 a M r^2 S + \mu M a^2 r^2-\mu r^5)u^\phi], \label{eq:Eq1td}  \\ 
&3c_{S^2} M S^2[a(p^t)^2-(2a^2+r^2)p^t p^\phi+a(a^2+r^2)(p^\phi)^2]= \nonumber\\ 
&\mu r^2 p^\phi [(a M S-\mu r^3)u^t+S (r^3-M a^2)u^\phi] \nonumber\\ 
&+\mu r^2 p^t[(a M S+\mu r^3)u^\phi-M S u^t].\label{eq:Eq2td}
\end{align} 
Notice that when $c_{S^2}=0$, Eqs.~\eqref{eq:Eq1td}, \eqref{eq:Eq2td} reduce to Eqs.~(19), (20) from \cite{Khodagholizadeh20}. In order to numerically solve the system of Eqs.~\eqref{eq:Eq1td}, \eqref{eq:Eq2td} it is convenient to introduce the quantity $W=p^\phi/p^t$, whereas the definition of the dynamical rest mass $\mu:=\sqrt{-p^\mu p_\mu}$ correlates the component $p^t$ with $W$.
\begin{equation}
    p^t=\dfrac{\mu}{\sqrt{-g_{tt}-2g_{t\phi}W-g_{\phi\phi}W^2}}.
\end{equation}
 We use the $\texttt{Solve}$ routine of $\texttt{Mathematica}$ in order to solve the non linear system of Eqs.~\eqref{eq:Eq1td}, \eqref{eq:Eq2td} for $\Omega>0$ and $W>0$, with $a$, $r$, $\hat{c}_{S^2}$, $S$ as fixed parameters. Firstly, to check our scheme we compare it with the frequency of the innermost stable circular orbit (ISCO) $\hat{\Omega}_{\rm ISCO}$ given in Table I of \cite{Bini15}. In particular, we solve numerically the system of Eqs.~\eqref{eq:Eq1td}, \eqref{eq:Eq2td} for selected values of $\hat{a}$, $\sigma$ and $\hat{c}_{S^2}$, while we adopt the $\hat{r}_{\rm ISCO}$ results from Table I of \cite{Bini15}. Two cases of $\hat{c}_{S^2}$ have been taken into consideration: $\hat{c}_{S^2}=1$, which corresponds to a black hole for the TD SSC, and $\hat{c}_{S^2}=6$ that describes a neutron star for the TD SSC \cite{Bini15,Vines2016}.  Notice that the orbital frequencies derived by the proposed method are in agreement up to third decimal place with the approximate $\mathcal{O}(S^3)$ analysis of Bini et al.~\cite{Bini15}. The discrepancy at order $\mathcal{O}(S^4)$  is expected, since the spin of the secondary has been set to $|\sigma|=0.1$ in the examples given in \cite{Bini15}. Another source of discrepancy could be the slightly different definition of the spin induced quadrupole in \cite{Bini15} contrasted to Eq.~\eqref{eq:SpinInd} of the present article. For a future possible comparison we also provide Table~\ref{tab:CEOTD}, while some further numerical examples are given in Appendix~\ref{sec:app2}.

\subsection{CEOS UNDER MP SSC}  \label{sec:CEOMP}

\begin{table}[h]
\centering
\normalsize
\begin{tabular}{|c|c|c|c|}
\hline
\multicolumn{4}{|c|}{$\hat{c}_{S^2}=1$}\\
\hline
\backslashbox{$\sigma$}{\\ $\hat{a}$}&$-0.9$ &0 &0.9\\
\hline
$-0.9$ & 0.03468 & 0.03320& 0.03185\\
$-0.1$ & 0.03276& 0.03178& 0.03085\\
$+0.1$ & 0.03235& 0.03148& 0.03065\\
$+0.9$ & 0.03098& 0.03049& 0.03001\\
\hline
\multicolumn{4}{|c|}{$\hat{c}_{S^2}=6$}\\
\hline
\backslashbox{$\sigma$}{\\  ${\hat a}$}&$-0.9$ &0 &0.9\\
\hline
$-0.9$ & 0.03565 & 0.03398 & 0.03252\\
$-0.1$  & 0.03277& 0.03179& 0.03086\\
$+0.1$  & 0.03236& 0.03149& 0.03066\\
$+0.9$  & 0.03192& 0.03122& 0.03063\\
\hline
\end{tabular}
\caption{ The Table depicts the orbital frequencies $\hat{\Omega}$ of a spinning body moving in a corrotating/counterotating circular equatorial orbit of $\hat{r}=10$, under the MP SSC.    
}
\label{tab:CEOMP}
\end{table}

Under the MP SSC the reference four-vector $V^\nu$ coincides with the four-velocity $u^\nu$ of the spinning body, i.e. $V^\nu:=u^\nu$. For this particular SSC the evolution equation of the kinematical rest mass $m$ reads \cite{Steinhoff:2012rw}

\begin{equation} \label{eq:mdot}
\dot{m}= \dfrac{\dot{R}_{\alpha \beta \gamma \delta}J^{\alpha \beta \gamma \delta}}{6}-\dfrac{4\dot{u}_\beta u_\gamma R^{[\beta}_{~\rho \kappa \lambda } J^{\gamma]\rho \kappa \lambda }}{3},
\end{equation}
whereas Eq.~\eqref{eq:Sdot} is still valid. It is worth noting that the CEOs scenario directly leads to the conservation of the kinematical rest mass and the spin measure, i.e.~$\dot{m}=0$, $\dot{S}=0$. Moreover, Eq.~\eqref{eq:momvelMP} guarantees proportionality  between the radial and polar components of the four-momentum and the four-velocity, or in other words $u^r \propto p^r$ and $u^\theta \propto p^\theta$. Following the CEO setup presented at the beginning of Sec.~\ref{sec:CEO} the two non trivial MPD equations for the MP case read 
\begin{align}
    &3 c_{S^2} M S^2[r^2+5a^2 (u^t)^2-2a(5a^2+4r^2)u^t u^\phi \label{eq:Eq1mp}\\
    &+(5a^4+8a^2 r^2+3r^4)(u^\phi)^2]=6 M S r^2 (2 a^2  \nonumber\\
    &+r^2)u^tu^\phi-6 a M S r^2 (u^t)^2-6 a M S r^2 (a^2 + r^2) (u^\phi)^2 \nonumber \\
    &-2 M r^4 p^t (u^t-a u^\phi)+2 r^4 p^\phi [a M u^t+(r^3-M a^2)u^\phi] \nonumber, \\
    &3 c_{S^2} M S^2 [a(u^t)^2-(2a^2+r^2)u^t u^\phi\label{eq:Eq2mp}\\
    &+a(a^2+r^2)(u^\phi)^2]=r^5 (p^t u^\phi-p^\phi u^t)-M S r^2(u^t)^2\nonumber \\
    &+2a M S r^2 u^t u^\phi+S r^2(r^3-M a^2)(u^\phi)^2 \nonumber.
\end{align}
The definition of the kinematical mass $m:=-p^\nu u_\nu$, combined with one of  Eqs.~\eqref{eq:Eq1mp}, \eqref{eq:Eq2mp}  can be used as a constraint in order to eliminate $p^\nu$. Furthermore, by employing Eq.~\eqref{eq:OrbFre}, along with the expression $\Omega=u^\phi/u^t$ leads to the construction of a quartic equation with respect to the orbital frequency as can be seen below

\begin{align}
   \zeta_4 \Omega^4+\zeta_3 \Omega^3+\zeta_2 \Omega^2+\zeta_1 \Omega+\zeta_0=0 \label{eq:quartic},
   \end{align}
where
\begin{align}
    \zeta_4 &=-2m r^{10}-2m a^2 r^8+2M r^7 (3  c_{S^2} S^2+6 a S-m a^2) \nonumber\\
    &+M a^2 r^5 (27 c_{S^2}S^2+ 2m a^2+ 14 a S)+2 a^2 M^2 r^4 (2 m a^2 \nonumber \\
    &+3a S-6  c_{S^2} S^2)+6 M S a^4  r^3 (a+6c_{S^2} S)+2 S M^2 a^4 r^2(5a \nonumber \\
    &+6c_{S^2} S) +3 M c_{S^2} S^2 a^4 r(5a^2-4 M^2)+24c_{S^2} S^2M^2 a^6,\nonumber\\
    \zeta_3 &=-2[-S r^8-2M r^7(m a-3S)+a M r^5(2 m a^2+12 a S  \nonumber\\
    &+15 c_{S^2} S^2)+a M^2 r^4(8 m a^2+9 a S-12 c_{S^2}S^2) \nonumber \\
    &+6 M S a^3 r^3(a+5 c_{S^2} S)+4 S M^2 a^3 r^2(5a+6 c_{S^2} S)  \nonumber \\
    &+3M c_{S^2} S^2 a^3 r(5 a^2-8 M^2)+48 c_{S^2} S^2 M^2 a^5] ,\nonumber \\
    \zeta_2 &=2m r^8-2 M m r^7+3 M S r^5(2a+ c_{S^2} S) \nonumber \\
    &+6 M^2 r^4 (4 m a^2+3 a S-2 c_{S^2} S^2)+9 M c_{S^2} S^2 a^2 r^3  \nonumber \\
    &+12 S M^2 a^2 r^2 (5a+6 c_{S^2} S) -72 c_{S^2} S^2 a^2 M^3 r\nonumber \\
    &+144 c_{S^2} S^2 M^2 a^4,
    \nonumber \\
    \zeta_1&=-2M[ -2 r^5 (m a + S)+M r^4(8 a m+ 3S) \nonumber \\
    &-3 a S r^3 (2a+ 3 c_{S^2} S) +4 a S M r^2(5a+ 6 c_{S^2} S)  \nonumber \\
    &- 3a c_{S^2} S^2 r(5 a^2 +8 M^2)+48 M c_{S^2} S^2 a^3],\nonumber \\
    \zeta_0 &=M[-2m r^5+4 M m r^4 -3 S r^3 (2 a+ c_{S^2} S) \nonumber \\
    &+2 M S r^2 (5a + 6c_{S^2} S) - 3 c_{S^2} S^2 r(5 a^2+ 4 M^2)\nonumber \\
    &+ 24 M c_{S^2} S^2 a^2]. \nonumber 
\end{align}

It is worth noticing that Eq.~\eqref{eq:quartic} reduces to Eq.~(27) of \cite{Timogiannis21} in the pole-dipole approximation limit, i.e., when $c_{S^2}=0$. Such an equation can be solved analytically in order to derive the orbital frequencies of an extended spinning body moving in the equatorial plane of a supermassive black hole. 
These solutions are provided in our work as a supplemental material in a \textit{Mathematica} notebook~\cite{SupMatMPQ}. By applying a similar criterion as in Refs.~\cite{Costa18,Timogiannis22}, that is, by using the limit $M\rightarrow 0$, $a\rightarrow 0$, we find that two of the four solutions are physical, since in the aforementioned limit we arrive at $\Omega \rightarrow 0$, and two non physical, since in this limit we  have $\Omega \neq 0$. In Refs.~\cite{Costa18,Timogiannis22} it has been shown using numerics that in the pole-dipole approximation the physical solutions led to non-helical orbits; in the pole-dipole-(spin induced) quadrupole we may postulate the same thing, i.e.~that the physical solutions are also the non-helical ones, since we retrieve the pole-dipole solutions in the limit $c_{S^2}=0$. Additionally, in Figs.~\ref{fig:cq1}, \ref{fig:cq6} included in Appendix~\ref{sec:app2} the curves for the TD and MP SSCs tend to coincide as $|\sigma|$ gets smaller. For a possible future comparison we provide some numerical results for the MP cases in Table~\ref{tab:CEOMP} as well. Note that each SSC appears to give a different value of $c_{S^2}$ for each type of object as a black hole or a neutron star \cite{Vines2016}. It is beyond the scope of this work to make the correspondence for the MP SSC between the $c_{S^2}$ and the type of object it describes.

\section{Conclusions} \label{sec:Conc}

The present article intently examines the motion of extended bodies in General Relativity. Namely, we shed light on the problem of finding circular equatorial orbits in Kerr spacetime, for a spinning body described by a pole-dipole-(spin induced) quadrupole approximation. The analysis is based on an algorithm developed in \cite{Khodagholizadeh20} and improved in \cite{Timogiannis21,Timogiannis22}. Two frequently employed spin supplementary conditions have been taken into consideration; the Tulczyjew-Dixon (TD) condition and the Mathisson-Pirani (MP) condition.  

Primarily, for an extended spinning test body governed by the TD SSC we derive a correlation between the dynamical rest mass $\mu$ and the kinematical rest mass $m$, which is in agreement with its pole-dipole counterpart \cite{Witzany}. In addition, it is shown that for the concept of circular equatorial orbits investigated here, $\mu$ and the spin magnitude $S$ are constants of motion. This fact is not generally valid within the pole-dipole-quadrupole regime. Furthermore, in Sec.~\ref{sec:CEO} we discuss the process of computing the orbital frequencies of a spinning body moving on a circular orbit on the equatorial plane of a supermassive Kerr black hole. It is worth noting that this is the first attempt for a non-approximative analytical formulation of such a problem in the literature. The results presented in Tab.~\ref{tab:ISCOBini} seem to be in accordance with the approximate $\mathcal{O}(S^3)$ method introduced in \cite{Bini15}. 

The notion of circular equatorial orbits of a pole-dipole-(spin induced) quadrupole test body under the imposal of the MP SSC, has been an uncharted territory for the community. In this study we employ the findings of \cite{Khodagholizadeh20,Timogiannis21,Timogiannis22} in order to calculate the orbital frequencies of a spinning body moving in the background of a Kerr black hole. This calculation culminates in obtaining the solutions of a quartic polynomial, according to Eq.~\eqref{eq:quartic}. In the procedure of achieving this very first analytical treatment of the problem hitherto, we show that the non-helical CEOs under the MP SSC are characterized by two conserved quantities; the kinematical rest mass of the test body $m$ and its spin magnitude $S$.

\begin{acknowledgments}
G.L.G. has been supported by the fellowship Lumina Quaeruntur No.~LQ100032102 of the Czech Academy of Sciences. We would also like to express our gratitude to Vojt\v{e}ch Witzany for the useful discussions. 
\end{acknowledgments}

\appendix
\section{Alternative Derivation Of A Momentum-Velocity Relation For The MP SSC} \label{sec:app1}
The main goal of Appendix~\ref{sec:app1} is the reproduction of Eq.~\eqref{eq:momvelMP} in the spirit of \cite{Costa18}. For that purpose, one starts by taking the covariant derivative of Eq.~\eqref{eq:SpinTensor}, with $V^\nu:=u^\nu$ and substitutes in Eq.~\eqref{eq:MPD_S} in order to get

\begin{equation} \label{eq:ap1}
    p_\mu u_\nu-p_\nu u_\mu+F_{\mu\nu}=\epsilon_{\mu\nu\alpha\beta}(\dot{u}^\alpha S^\beta+u^\alpha \dot{S}^\beta).
\end{equation}
The contraction of Eq.~\eqref{eq:ap1} with the quantity $\epsilon^{\kappa\lambda\mu\nu} S_\lambda$ leads to an expression for the four-acceleration, generalized in the pole-dipole-quadrupole case 

\begin{equation} \label{eq:ap2}
   \dot{u}^\nu=-\dfrac{2\dot{S}_\mu S^\nu u^\mu+2p_\alpha S^{\nu\alpha}+\epsilon^{\nu\lambda\alpha\beta}S_\lambda F_{\alpha\beta}}{2S^2}. 
\end{equation}
In addition, the contraction of Eq.~\eqref{eq:MPD_S} with the four-velocity $u_\mu$ yields
\begin{equation} \label{eq:ap3}
    p^\nu=m u^\nu-\dot{u}_\mu S^{\mu\nu}-u_\mu F^{\mu\nu},
\end{equation}
where the relation $u_\mu \dot{S}^{\mu\nu}=-\dot{u}_\mu S^{\mu\nu}$ (only valid for the MP SSC) has been taken into account. The combination of Eqs.~\eqref{eq:ap2}, \eqref{eq:ap3} gives
\begin{equation} \label{eq:ap4}
    m u^\nu=p^\nu+u_\mu F^{\mu\nu}+\dfrac{S_{\mu\alpha}S^{\nu\mu}p^\alpha}{S^2}-\dfrac{\epsilon_{\mu\kappa\alpha\beta}S^\kappa S^{\mu\nu}F^{\alpha\beta}}{2S^2}.
\end{equation}
Note that if one takes advantage of the definition of the spin vector Eq.~\eqref{eq:SpinVect}, Eq.~\eqref{eq:ap4} can be re-expressed in terms of the spin tensor
\begin{equation} \label{eq:ap5}
   mu^\nu=p^\nu+ \dfrac{S_{\mu\alpha}S^{\nu\mu}p^\alpha}{S^2}+\dfrac{u_\alpha S_\beta S^\nu F^{\alpha\beta}}{S^2}.
\end{equation}
For the last step of the calculation Eq.~\eqref{eq:important} is implemented in Eq.~\eqref{eq:ap5}, which leads to Eq.~\eqref{eq:momvelMP}. 

\section{Numerical Examples} \label{sec:app2}

\begin{figure}[h]
 \graphicspath{{./PhD/}}
  \centering
    \includegraphics[width=.85\linewidth]{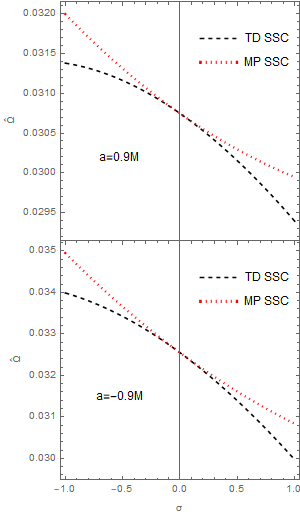}
\caption{The top panel depicts the shift of the orbital frequency of a spinning test body ($\hat{c}_{S^2}=1$) due to the presence of its spin, computed in dimensionless units under the TD and MP conditions, for a CEO of $\hat{r}=10$, when $\hat{a}=0.9$. The bottom panel represents the orbital frequency of a spinning test body ($\hat{c}_{S^2}=1$) with respect to the spin, for a CEO of $\hat{r}=10$, when $\hat{a}=-0.9$, under the TD and MP formalisms.}
\label{fig:cq1}
\end{figure}
\begin{figure}[h]
   \graphicspath{{./PhD/}}
  \centering
    \includegraphics[width=.85\linewidth]{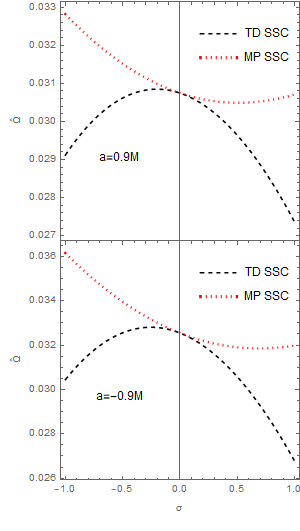}
    \caption{The top panel demonstrates the shift of the orbital frequency of a spinning test body ($\hat{c}_{S^2}=6$) due to the presence of its spin, computed in dimensionless units under the TD and MP conditions, for a CEO at $\hat{r}=10$, when $\hat{a}=0.9$. The bottom panel illustrates the orbital frequency of a spinning test body ($\hat{c}_{S^2}=6$) with respect to the spin, for a CEO of $\hat{r}=10$, when $\hat{a}=-0.9$, under the TD and MP formalisms. }
    \label{fig:cq6}
 \end{figure}   

In Appendix~\ref{sec:app2} we make a detailed discussion on our numerical findings, which in Secs.~\ref{sec:CEOTD} and \ref{sec:CEOMP} were briefed into Tables~\ref{tab:ISCOBini}--\ref{tab:CEOMP}. The initial motivation has been to show that the results presented in the aforementioned Tables are sound and not numerical artifacts. We expect that curves demonstrating the orbital frequency of a CEO as function of the spin measure $\sigma$ for fixed radius should be smooth without discontinuities. Figs.~\ref{fig:cq1}--\ref{fig:cq6} illustrate such curves for the radius $\hat{r}=10$ of a CEO achieved under the TD  SSC (dashed black) and the MP SSC (dotted red). In Fig.~\ref{fig:cq1} we set $\hat{c}_{S^2}=1$, while in Fig.~\ref{fig:cq6} we use the value $\hat{c}_{S^2}=6$. From the smooth behavior of the plots, we deduce that the results of the aforementioned tables are sound. 

An interesting outcome from these plots, when contrasted with the corresponding pole-dipole results, is that the frequency curves produced under the different formalisms (TD SSC and MP SSC) appear to have different curvatures. Namely, in our previous work \cite{Timogiannis22} we have concluded that within the limits of the pole-dipole approximation for given radius $\hat{r}$, CEOs under the TD and MP SSCs agree up to order $\mathcal{O}(\sigma^2)$ in the orbital frequency expansions. Since the absolute value of the higher order coefficients in the expansion in $\sigma$ become smaller as the order increases for a pole-dipole body \cite{Timogiannis21,Timogiannis22}, the change in the curvature shown in Figs.~\ref{fig:cq1}-\ref{fig:cq6} indicates that the frequency agreement between the TD and MP SSCs in the pole-dipole-(spin induced) quadrupole case should be smaller than in the pole-dipole case, i.e., it should be less than $\mathcal{O}(\sigma^2)$ without the proper correction in position and spin. However, this discrepancy might simply showing that the same value of $\hat{c}_{S^2}$ for the TD SSC and the MP SSC does not correspond to the same physical object. A study of this issue is left for a future work.

Another interesting effect reflected in Figs.~\ref{fig:cq1}--\ref{fig:cq6} is that the monotonicity appearing in the frequency curves for $\hat{c}_{S^2}=1$ is broken in the $\hat{c}_{S^2}=6$ case. This designates that the behavior of the curvature of the frequency curve is dominated by the spin induced quadrupole term in this approximation of the extended body.

\bibliographystyle{unsrt}
\bibliography{refs}

\end{document}